\documentclass[aps,prl,twocolumn,amsmath,amssymb,superscriptaddress]{revtex4-1}
\usepackage{graphicx}  
\usepackage{dcolumn}   
\usepackage{bm}        
\usepackage{verbatim}   
\usepackage{braket}
\usepackage{hyperref}
\usepackage{color}
\usepackage{xspace}
\newcolumntype{P}[1]{>{\centering\arraybackslash}p{#1}}
\newcolumntype{M}[1]{>{\centering\arraybackslash}m{#1}}
\newcommand*{\melvin}{{\small M}{\scriptsize ELVIN}\xspace}
\newcommand{\brac}[2]{\ensuremath{\left< #1\left|#2 \right. \right>}}
\newcommand{\abs}[1]{\ensuremath{\left| #1 \right|}}
\begin{document}
\bibliographystyle{plainnat} 
\title{High-Dimensional Single-Photon Quantum Gates: Concepts and Experiments}

\author{Amin Babazadeh}
\affiliation{Vienna Center for Quantum Science \& Technology (VCQ), Faculty of Physics, University of Vienna, Boltzmanngasse 5, 1090 Vienna, Austria.}
\affiliation{Institute for Quantum Optics and Quantum Information (IQOQI), Austrian Academy of Sciences, Boltzmanngasse 3, 1090 Vienna, Austria.}
\affiliation{Physics Department, Institute for Advanced Studies in Basic Sciences, Zanjan, Iran.}
\author{Manuel Erhard}
\email{manuel.erhard@univie.ac.at}
\affiliation{Vienna Center for Quantum Science \& Technology (VCQ), Faculty of Physics, University of Vienna, Boltzmanngasse 5, 1090 Vienna, Austria.}
\affiliation{Institute for Quantum Optics and Quantum Information (IQOQI), Austrian Academy of Sciences, Boltzmanngasse 3, 1090 Vienna, Austria.}
\author{Feiran Wang}
\affiliation{Vienna Center for Quantum Science \& Technology (VCQ), Faculty of Physics, University of Vienna, Boltzmanngasse 5, 1090 Vienna, Austria.}
\affiliation{Institute for Quantum Optics and Quantum Information (IQOQI), Austrian Academy of Sciences, Boltzmanngasse 3, 1090 Vienna, Austria.}
\affiliation{Key Laboratory of Quantum Information and Quantum Optoelectronic Devices, Shaanxi Province, Xi�an Jiaotong University, Xi�an 710049, China.}
\author{Mehul Malik}
\affiliation{Vienna Center for Quantum Science \& Technology (VCQ), Faculty of Physics, University of Vienna, Boltzmanngasse 5, 1090 Vienna, Austria.}
\affiliation{Institute for Quantum Optics and Quantum Information (IQOQI), Austrian Academy of Sciences, Boltzmanngasse 3, 1090 Vienna, Austria.}
\author{Rahman Nouroozi}
\affiliation{Physics Department, Institute for Advanced Studies in Basic Sciences, Zanjan, Iran.}
\author{Mario Krenn}
\affiliation{Vienna Center for Quantum Science \& Technology (VCQ), Faculty of Physics, University of Vienna, Boltzmanngasse 5, 1090 Vienna, Austria.}
\affiliation{Institute for Quantum Optics and Quantum Information (IQOQI), Austrian Academy of Sciences, Boltzmanngasse 3, 1090 Vienna, Austria.}
\author{Anton Zeilinger}
\email{anton.zeilinger@univie.ac.at}
\affiliation{Vienna Center for Quantum Science \& Technology (VCQ), Faculty of Physics, University of Vienna, Boltzmanngasse 5, 1090 Vienna, Austria.}
\affiliation{Institute for Quantum Optics and Quantum Information (IQOQI), Austrian Academy of Sciences, Boltzmanngasse 3, 1090 Vienna, Austria.}

\date{\today}
\begin{abstract}

Transformations on quantum states form a basic building block of every quantum information system. From photonic polarization to two-level atoms, complete sets of quantum gates for a variety of qubit systems are well known. For multi-level quantum systems beyond qubits, the situation is more challenging. The orbital angular momentum modes of photons comprise one such high-dimensional system for which generation and measurement techniques are well-studied. However, arbitrary transformations for such quantum states are not known. Here we experimentally demonstrate a four-dimensional generalization of the Pauli X-gate and all of its integer powers on single photons carrying orbital angular momentum. Together with the well-known Z-gate, this forms the first complete set of high-dimensional quantum gates implemented experimentally. The concept of the X-gate is based on independent access to quantum states with different parities and can thus be easily generalized to other photonic degrees-of-freedom, as well as to other quantum systems such as ions and superconducting circuits.
\end{abstract}

\maketitle

\textit{Introduction} -- High-dimensional quantum states have recently attracted increasing attention in both fundamental and applied research in quantum mechanics \cite{agnew2011tomography, romero2013tailored, krenn2014generation, zhang2016engineering, malik2016multi}. The possibility of encoding vast amounts of information on a single photon makes them particularly interesting for large-alphabet quantum communication protocols \cite{groblacher2006experimental, sit2016high, Smania2016, Lee:2016vk}, as well as for investigating fundamental questions concerning local realism or quantum contextuality \cite{vaziri2002experimental, cai2016new}. The temporal and spatial structure of a photon provides a natural multi-mode state space in which to encode quantum information. The orbital angular momentum (OAM) modes of light \cite{allen1992orbital} comprise one such basis of spatial modes that has emerged as a popular choice for experiments on high-dimensional quantum information \cite{krenn2017orbital}. While techniques for the generation and measurement of photonic quDits carrying OAM are well known \cite{mair2001entanglement, Mirhosseini:2013go, malik2014direct}, efficient methods for their control and transformation remain a challenge. No general recipe is known so far, and experimentally feasible techniques are known only for special cases. 

Here we experimentally demonstrate a four-dimensional X-gate and all of its integer powers with the orbital angular momentum modes of single photons. The four-dimensional X-gate is a generalization of the two-dimensional $\sigma_x$ Pauli transformation and acts as a cyclic ladder operator on a four-dimensional Hilbert space. The cyclic transformation required for this gate was designed through the use of the computer algorithm \melvin \cite{krenn2016automated} and recently demonstrated with classical states of light \cite{schlederer2016cyclic}. The Z-gate for OAM quDits (the generalization of the two-dimensional $\sigma_z$ Pauli transformation) introduces a mode-dependent phase, which can be implemented simply with a single optical element~\cite{leach2002measuring,de2005implementing}. With all powers of the high-dimensional X- and Z-gate, we arrive at a complete basis of quDit gates, which in principle allows for the construction of arbitrary unitary operations in a four-dimensional state space \cite{asadian2016heisenberg} (see Appendix for details).

It is interesting to compare OAM with other high-dimensional degrees of freedom that allow for the encoding of quantum information. For path-encoding in particular, it is known how arbitrary single-quDit transformations can be performed in a lossless way \cite{reck1994experimental}. Such transformations have been implemented recently on integrated photonic chips for the generation and transformation of entanglement \cite{schaeff2015experimental, carolan2015universal}. General unitary transformations such as these are not known for the photonic OAM degree-of-freedom. In addition to being natural modes in optical communication systems with cylindrical symmetry, photons carrying OAM offer an important advantage over path and time-bin encoding in that quantum entanglement can be generated \cite{romero2012increasing} and transmitted \cite{krenn2015twisted} without the need for interferometric stability.  Therefore, the development of new controlled transformations for photonic OAM, as we show here, fills an important gap. 

The X-gate demonstrated here uses the ability to sort even and odd parity modes as a basic building block ~\cite{leach2002measuring}. This concept can be extended to other photonic degrees of freedom such as frequency \cite{yokoyama2013ultra, xie2015harnessing}, and used in other quantum systems such as trapped ions \cite{muthukrishnan2000multivalued, klimov2003qutrit}, cold atoms \cite{smith2013quantum}, and superconducting circuits \cite{hofheinz2009synthesizing} for constructing similar high-dimensional quantum logic gates.

\textit{High-dimensional Pauli gates} -- The Pauli matrix group has applications in quantum computation, quantum teleportation and other quantum protocols. This group is defined for a single quDit (a single photon with d-dimensional modes) in the following manner \cite{gottesman1999fault,lawrence2004mutually}:  
\begin{subequations}
\begin{align}
&X=\sum_{\ell=0}^{d-1}\ket{\ell\oplus1}\bra{\ell}\label{Eq:X-gate}\\
&Z=\sum_{\ell=0}^{d-1}\ket{\ell}\omega^{\ell}\bra{\ell}\label{Eq:Z-gate}
\end{align}
\end{subequations}  
where $\ell\in\{0,1...,d-1\}$ refers to the different modes in the d-dimensional Hilbert space and $\ell\oplus1 \equiv(\ell+1)$ mod $d$. The Z-gate introduces a mode-dependent phase in the form of $\omega=\exp({2\pi i\over d})$. Furthermore, the Y-gate can be written $Y=X\cdot Z$. While the two-dimensional X-gate swaps two modes with one another, in high dimensional Hilbert spaces ($d>2$) it takes the form of a cyclic operation:
\begin{equation}
      X\ket{\ell}=\ket{\ell+1}. 
\end{equation}
This results in each state being transformed to its nearest neighbor in a clockwise direction, with the last state $\ket{d-1}$ being transformed back to the first one $\ket{0}$. The Y-gate can be expressed as a combination of Z and X gates. While powers of $Z$ lead to different mode-dependent phases, integer powers of X shift the modes by a larger number:  
\begin{equation}
       X^n = \sum_{\ell=0}^{d-1}\ket{\ell\oplus n}\bra{\ell}
       \label{Eq:S-gate}
\end{equation}
The X$^2$-gate, for example, transforms each mode to the second nearest mode. Likewise, the conjugate of X leads to a cyclic operation in the counter-clockwise direction,
\begin{equation}
       X^{\dagger} =\sum_{\ell=0}^{d-1}\ket{\ell\ominus1}\bra{\ell}.
\end{equation}

\begin{figure}
\includegraphics[scale=0.43]{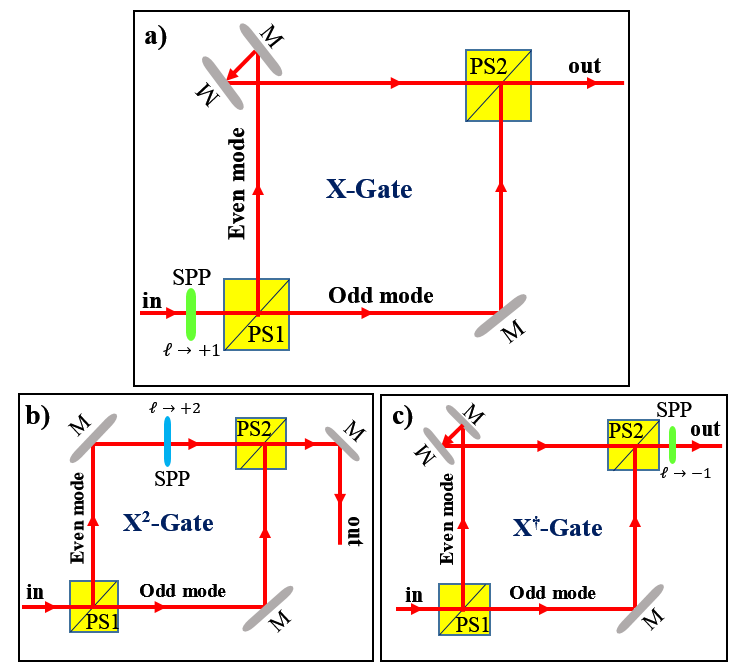}
\caption{\label{figure:schem} The conceptual diagrams for the three types of quantum logic gates. The input states for each case is (-2,-1,0,1). a) A spiral phase plate (SPP) adds $+1$ to the mode, leading to (-1,0,1,2). Afterwards, the first parity sorter separates even and odd modes, while the second one combines them again -- which forms a large interferometer. Within the interferometer, the even mode have an odd number of reflections, which leads to the correct output modes (-1,0,1,-2). b) For the $X^2$ gate, the input modes are directly separated into even and odd modes. After a reflection in each arm, the even modes are increased by 2. The two arms are recombined at the PS2 and all of the modes are reflected for changing the signs of the modes. That leads to (0,1,-2,-1). c) In the X$^{\dagger}$ transformation, the different parity modes are separated again, and the even part gets reflected twice, while the odd modes are reflected once. After recombination, the $\ell$ of the modes is decreased by one, which leads to (1,-2,-1,0). Note that in the experiment, it can be adjusted whether even or odd modes are reflected at the parity sorter. In this conceptual diagrams, for simplicity we have chosen PS1 to reflect even modes and PS2 to reflect odd modes.}
\end{figure}

\textit{Experimental implementation} -- A Z-gate for photons carrying OAM can simply be achieved by using a Dove prism, which has been shown recently \cite{agnew2013generation, wang2015quantum, zhang2016engineering, ionicioiu2016sorting}. Since the Y-gate can be achieved by a combination of Z and X gates, it is sufficient to focus on the X-gate and its powers. Fig.\ref{figure:schem}a shows the schematic of the X-gate. It consists of two parity sorters (PS1 and PS2) and a Mach-Zehnder interferometer (MZI) that is implemented between them. The input photon is first incident on a spiral phase plate  that adds one quantum of OAM quantum ($\text{SPP}_{\ell+1}$) onto the photon before it enters PS1. The parity sorter is an interferometric device which then sorts the photon according to its mode parity \cite{leach2002measuring}. For the first-order cyclic transformation, the sign of the odd photon needs to be flipped after PS1. This is achieved by reflecting the odd output photon at a mirror placed in one MZI path, while even photons undergo two reflections that preserve the sign of their OAM mode (see Fig.\ref{figure:schem}a). The modes are then input into PS2, which coherently recombines them into the same path.
 
Interestingly, the X-gate can be converted into the X$^2$-gate and X$^{\dagger}$-gate with only minor changes to the experimental setup (for d=4, X$^{\dagger}$=X$^3$). For constructing the X$^2$-gate, the $\text{SPP}_{\ell+1}$ is removed and an $\text{SPP}_{\ell+2}$ replaces the extra reflection in the even MZI path (Fig.\ref{figure:schem}b). The X$^{\dagger}$-gate is achieved by simply moving the $\text{SPP}_{\ell+1}$ from the input of PS1 and replacing it with an $\text{SPP}_{\ell-1}$ at the output of PS2 (Fig.\ref{figure:schem}c). One should note that in principle, these changes can be implemented rapidly and without physically moving optical components via the use of devices such as a spatial light modulator or a digital micromirror device.

\begin{figure}
\includegraphics[scale=0.08]{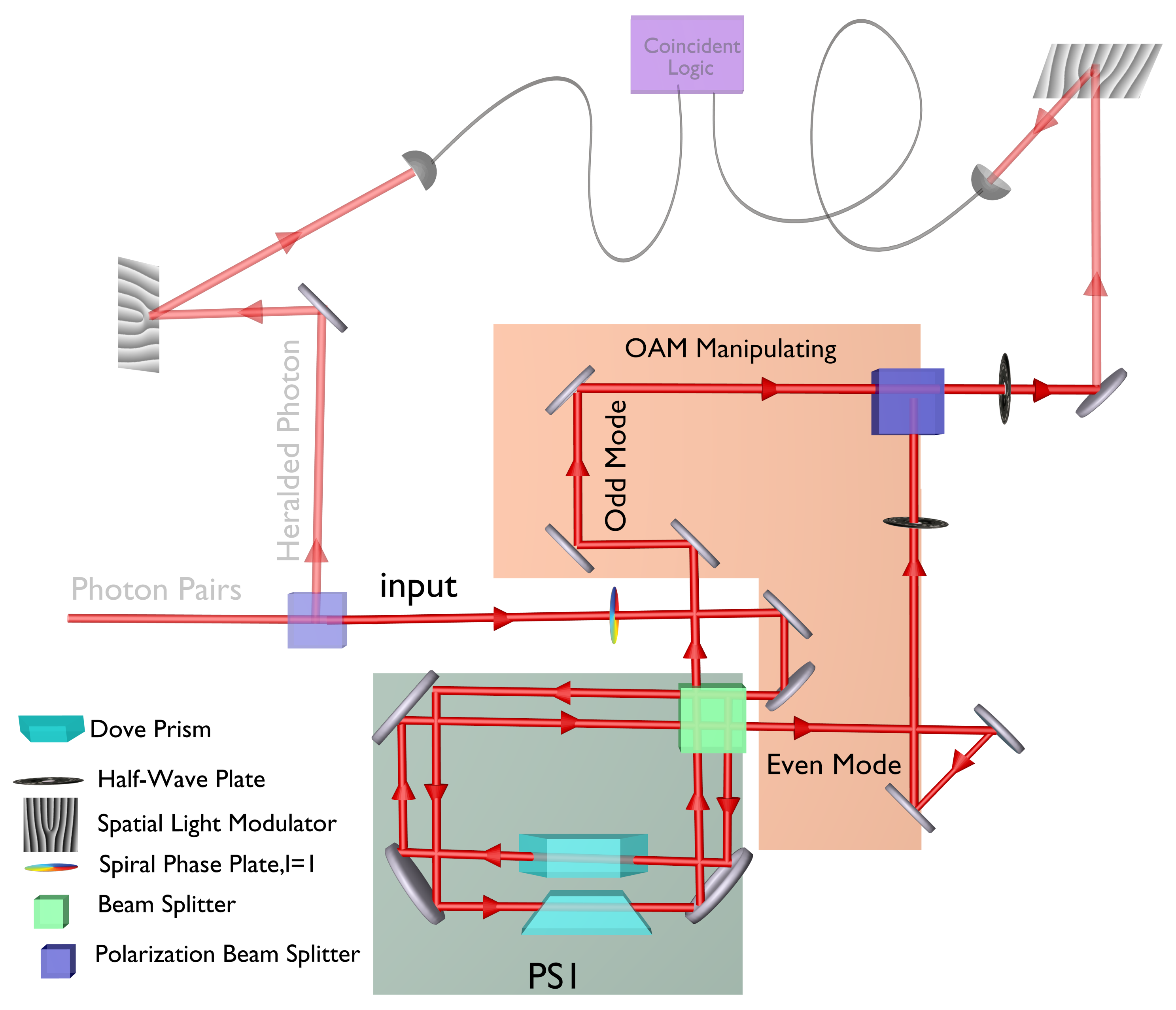}
\caption{\label{figure:setup} Experimental setup for the four-dimensional X-gate (additional experimental details are partially transparent). A 405nm CW laser pumps a Type-II ppKTP crystal (not shown), creating photon pairs entangled in orbital angular momentum (OAM). The idler photon is used for heralding the signal photon in a particular OAM mode. After passing through a $\ell=+1$ spiral phase plate, the signal photon is input into a parity sorting interferometer, which separates the odd and even OAM components of the photon. After traversing a series of mirrors, the odd and even components are coherently recombined in a Mach-Zehnder interferometer through the use of a half-wave plate (HWP) and polarizing beam splitter (PBS). A spatial light modulator (SLM) and single mode fiber are used to perform projective measurements of OAM modes and their superpositions.}
\end{figure}

The experimental setup is depicted in Fig.\ref{figure:setup}. We use heralded single photons produced via the process of Type-II spontaneous parametric down conversion process (SPDC) in a 5mm long periodically poled Potassium Titanyl Phosphate (ppKTP) crystal pumped by a 405nm diode laser. In the SPDC process, conservation of the pump angular momentum leads to the generation of photon pairs with a degenerate wavelength of $\lambda$=810nm that are entangled in OAM. Therefore whenever the idler photon is measured to be in mode $\ket{+\ell}$, the signal photon is found to be in mode $\ket{-\ell}$. Thus, by heralding the idler photon in a particular OAM mode, we can select the OAM quantum number of the signal photon that is input into the logic gate. Here, we use the OAM quantum numbers of -2,-1,0 and 1 for demonstrating our 4-dimensional quantum logic gates. By changing the mode number before and after the transformation, the X-gate can be used with every connected 4-dimensional subspace.

The parity sorter was originally proposed as an MZI with a dove prism in each arm \cite{leach2002measuring}. The relative rotation angle between the two dove prisms is set at $90^\circ$, which introduces an $\ell\pi$ phase difference between the two arms. Depending on the parity of OAM mode ($\ell$) of the input photon, constructive or destructive interference results in even and odd modes exiting different outputs of the MZI. For long-term stability, in our case we implement this interferometer in a double-path Sagnac configuration~\cite{10.1088/2058-9565/aa5917}. Two adjacent Sagnac loops allow for the positioning of a dove prism in each loop. The outputs of this Sagnac interferometer are then directly input into the second MZI (denoted as \textit{OAM manipulating} in Fig.\ref{figure:setup}). In the second interferometer the sign of odd modes is flipped by reflection on an extra mirror. A trombone system in the odd arm is used to adjust the relative path difference to achieve a coherent combination of even and odd modes.

The concept of the quantum gates (discussed in Fig.~\ref{figure:schem}) allows in principal for a lossless operation. For simplicity, we replace the second parity sorter with a polarization beam splitter (PBS). This allows the odd and even modes in the MZI to be recombined in a stable manner, albeit with an additional loss of 50\%. 

\begin{figure}
\includegraphics[scale=.2]{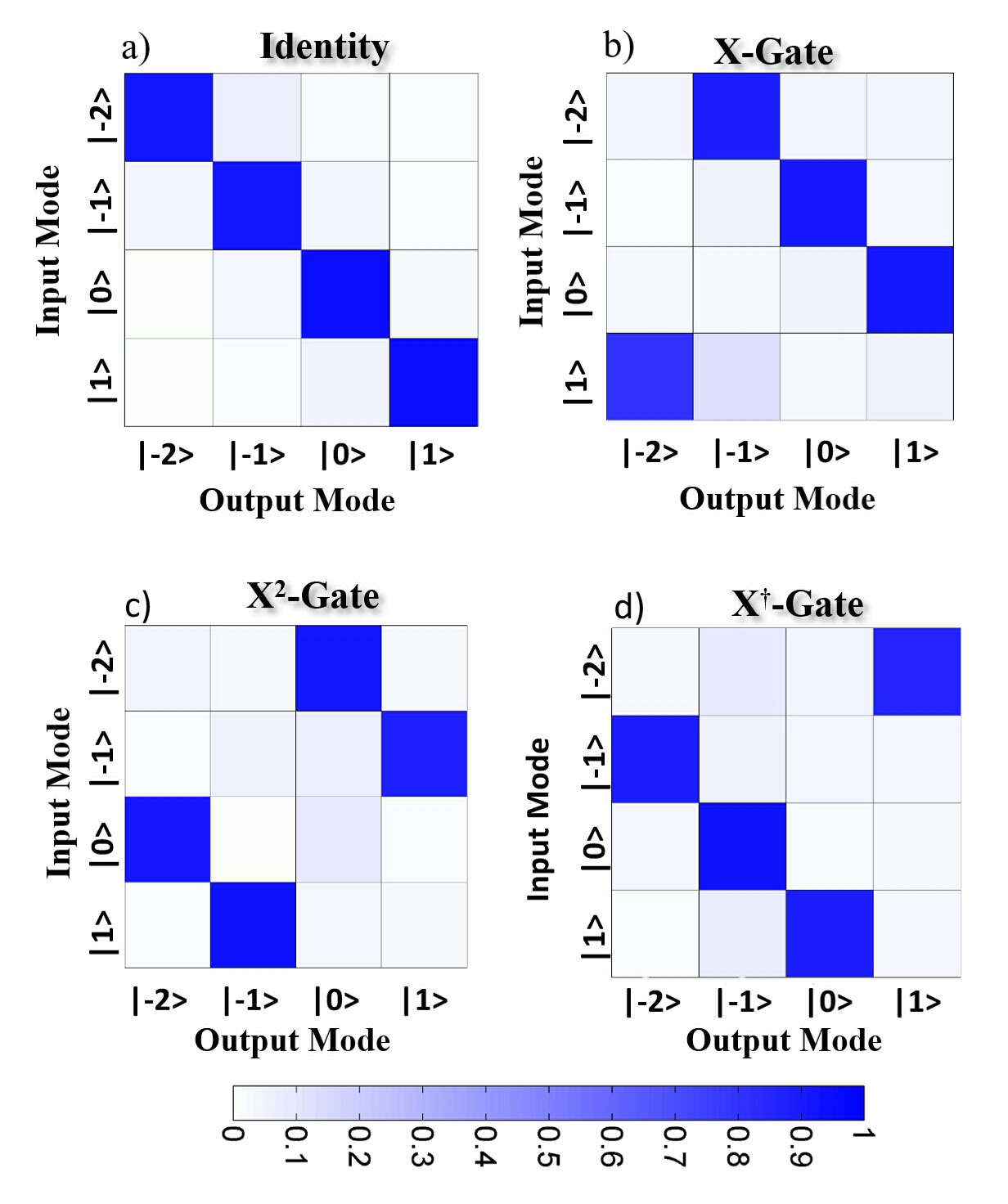}
\caption{\label{figure:correlation} Data showing the operation of the a) Identity, b) X-gate, c) X$^2$-gate, and d) X$^{\dagger}$-gate on the four-dimensional set of input states $\{\ket{-2},\ket{-1},\ket{0},\ket{1}\}$. Each row shows the measured normalised coincidence rate in every output mode for a given input mode. The X-gate implements a clockwise cyclic transformation $(-2\rightarrow -1\rightarrow0\rightarrow1\rightarrow-2)$, the X$^2$-gate swaps the odd and even modes $(-1\leftrightarrow 1,-2\leftrightarrow0)$, and the X$^{\dagger}$-gate performs a counter-clockwise cyclic transformation $(1\leftarrow-2\leftarrow-1\leftarrow0\leftarrow1)$. The average transformation efficiency for the X, X$^2$, and X$^\dagger$ gates are 87.3\%, 90.4\%, and 88.4\% respectively.} 
\end{figure}

Now we explain the experimental details of the X-Gate (Fig.~\ref{figure:schem}\&\ref{figure:setup}). A 4-dimensional subset of OAM modes $\ell\in\{-2,-1,0,1\}$ is shifted by one leading to $\{-1,0,1,2\}$. The parity sorter separates even and odd modes. The path for the even modes experience an odd number of reflections that causes in a sign flip and results in $\{-1,0,1,-2\}$. The coherent combination at the PBS and subsequent erasure of polarization information completes the X-gate: $(-2\rightarrow -1\rightarrow0\rightarrow1\rightarrow-2)$. The $\text{X}^2$ and $\text{X}^\dagger$ gate work similarly, see Fig.~\ref{figure:schem}. The experimental results of the gate operations are depicted in Fig.~\ref{figure:correlation}. The probability P$_{i,j}$ to detect a photon in mode $j$ when sending in one in mode $i$ is given by $P(i,j)=\frac{\abs{\brac{j_{\text{out}}}{i_{\text{in}}}}^2}{\sum_n{\abs{\brac{n_{\text{out}}}{i_{\text{in}}}}^2}}$. The average probability of the expected mode for the X, X$^2$, and X$^\dagger$ gates are 87.3\%, 90.4\%, and 88.4\% respectively, see Table~\ref{tab:efficiency}.

\begin{table}[t!]
\begin{center}
\caption {Transformation efficiency $E_{\ell_i}$ for each input state $\ket{\ell_i}$. The efficiency is calculated by dividing the number of photons in the correct output state by the total number of counts measured in all four states.}
\label{tab:efficiency} 
 \begin{tabular*}{0.47\textwidth}{@{\extracolsep{\fill}} | M{.7cm} | c | c |c| c | }
    
  \hline
  $\shortstack{Input\\mode}$ & $\ket{-2}$ &  $\ket{-1}$ &  $\ket{0}$ &  $\ket{1}$\\ 
  \hline  
  X-Gate & $88.1\pm2.2\%$ &  $90.3\pm1.6\%$ &  $90.9\pm0.3\%$ &  $80.1\pm0.6\%$\\ 
 \hline
  X$^2$-Gate & $90.8\pm1.1\%$ &  $87.1\pm1.2\%$ &  $90.3\pm1.7\%$ & $93.4\pm1.3\%$\\ 
  \hline
  X$^{\dagger}$-Gate & $85.1\pm1.2\%$ & $87.6\pm0.8\%$ &  $92.6\pm1.5\%$ &  $88.4\pm1.4\%$\\  
 \hline
\end{tabular*}
\end{center}

\end{table}
 
\begin{figure}[htb]
\includegraphics[scale=0.2]{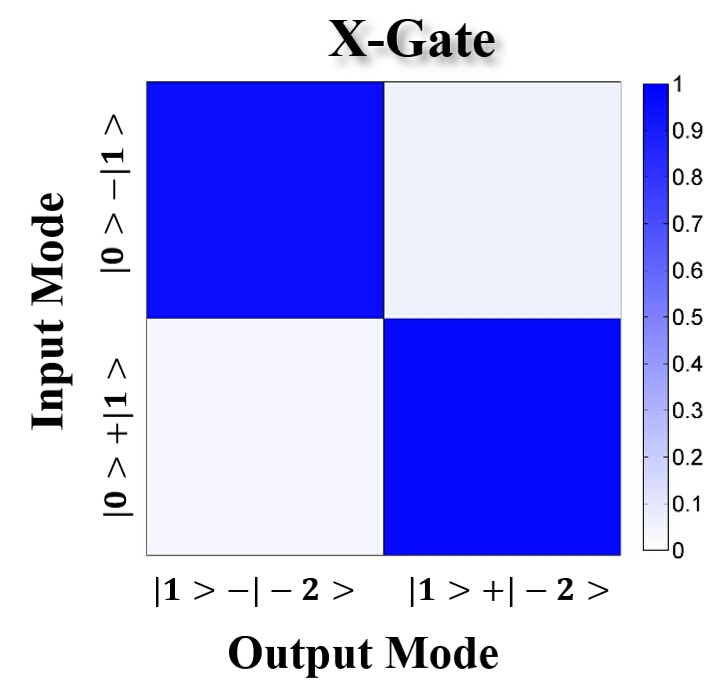}
\caption{\label{figure:coherence} Action of the X-gate on coherent superpositions of quantum states. It shows that the operation conserves the phases. The correlation matrix confirms that the input quantum states $\ket{\psi_{i}}=\frac{\ket{0}\pm\ket{1}}{\sqrt{2}}$ are transformed coherently to the output quantum states ($\ket{\phi_{o}}=\frac{\ket{1}\pm\ket{-2}}{\sqrt{2}}$). The visibility of the transformation process is calculated to be 90.9$\%$.}  
\end{figure}

In order to demonstrate a transformation of a coherent superposition, we use $\ket{\psi_{\text{in}}}=(\ket{0}\pm\ket{1})/\sqrt{2}$ as the input into the X-gate. The expected output state is given by $\ket{\phi_{\text{out}}}=(\ket{1}\pm\ket{-2})/\sqrt{2}$ and ideally we do not expect any photon in the orthogonal state $\ket{\bar{\phi}_{\text{out}}}=(\ket{1}\mp\ket{-2})/\sqrt{2}$. The probability to detect photons in these output modes is shown in Fig.~\ref{figure:coherence}. The average probability of detecting the expected mode is 90.9$\%$ and matches with the probabilities shown in Table~\ref{tab:efficiency} and Fig.~\ref{figure:correlation}. This clearly demonstrates that the X-gate works as a coherent quantum transformation. 

\textit{Conclusion} -- We have shown the experimental generation of the four-dimensional X-gate and all of its unique higher orders, including the X$^2$ and X$^3$ gates. Together with the well known Z-gate, this forms a complete basis of transformations on a four-dimensional quantum system. This means that it can in principle be used to construct every four-dimensional unitary operation. The X-gate is a basic element required for generating large classes of entangled states, such as the set of four-dimensional Bell states \cite{Feiran} or general high-dimensional multi-particle states \cite{malik2016multi}. Such states can be used, for example, in tests of quantum contextuality \cite{lapkiewicz2011experimental} and for Bell-like tests of local-realism in a higher-dimensional state space \cite{vaziri2002experimental, dada2011experimental}. 

These quantum logic gates can find application in various high-dimensional quantum protocols, such as high-dimensional quantum key distribution \cite{groblacher2006experimental, mafu2013higher, mirhosseini2015high, sit2016high} where transformations between mutually unbiased bases are necessary. Other applications could include multi-party secret sharing \cite{Smania2016} or dense coding \cite{hill2016hyperdense}, where transformations between orthogonal sets of entangled states are required. In quantum computing where complete sets of quantum gates are necessary, high-dimensional quantum states allow for the efficient implementation of gates \cite{ralph2007efficient, lanyon2009simplifying} and offer advantages in quantum error correction \cite{bocharov2016factoring}.

Interestingly, a high-dimensional generalization of the CNOT gate consists of a controlled-cyclic transformation \cite{garcia2013swap}. In combination with polarization, one can immediately create a three-, six- and  eight-dimensional generalization of our method \cite{krenn2016automated}. An important next step is the construction of high-dimensional two-particle gates. This would allow the implementation of complex quantum algorithms such as quantum error correction in high dimensions \cite{bocharov2016factoring}.

\section*{Acknowlegdements}
The authors thank Marcus Huber for helpful discussions. This work was supported by the Austrian Academy of Sciences (\"OAW), by the European Research Council (SIQS Grant No. 600645 EU-FP7-ICT) and the Austrian Science Fund (FWF) with SFB F40 (FOQUS) and FWF project CoQuS No. W1210-N16. F.W. was supported by the National Natural Science Foundation of China (NSFC Grant No. 11534008).

\bibliographystyle{unsrt}
\bibliography{refs}

\begin{thebibliography}{10}

\bibitem{agnew2011tomography}
Megan Agnew, Jonathan Leach, Melanie McLaren, F~Stef Roux, and Robert~W Boyd.
\newblock Tomography of the quantum state of photons entangled in high
  dimensions.
\newblock {\em Physical Review A}, 84(6):062101, 2011.

\bibitem{romero2013tailored}
Jacquiline Romero, Daniel Giovannini, DS~Tasca, SM~Barnett, and MJ~Padgett.
\newblock Tailored two-photon correlation and fair-sampling: a cautionary tale.
\newblock {\em New Journal of Physics}, 15(8):083047, 2013.

\bibitem{krenn2014generation}
Mario Krenn, Marcus Huber, Robert Fickler, Radek Lapkiewicz, Sven Ramelow, and
  Anton Zeilinger.
\newblock Generation and confirmation of a (100$\times$ 100)-dimensional
  entangled quantum system.
\newblock {\em Proceedings of the National Academy of Sciences},
  111(17):6243--6247, 2014.

\bibitem{zhang2016engineering}
Yingwen Zhang, Filippus~S Roux, Thomas Konrad, Megan Agnew, Jonathan Leach, and
  Andrew Forbes.
\newblock Engineering two-photon high-dimensional states through quantum
  interference.
\newblock {\em Science advances}, 2(2):e1501165, 2016.

\bibitem{malik2016multi}
Mehul Malik, Manuel Erhard, Marcus Huber, Mario Krenn, Robert Fickler, and
  Anton Zeilinger.
\newblock Multi-photon entanglement in high dimensions.
\newblock {\em Nature Photonics}, 10(4):248--252, 2016.

\bibitem{groblacher2006experimental}
Simon Gr{\"o}blacher, Thomas Jennewein, Alipasha Vaziri, Gregor Weihs, and
  Anton Zeilinger.
\newblock Experimental quantum cryptography with qutrits.
\newblock {\em New Journal of Physics}, 8(5):75, 2006.

\bibitem{sit2016high}
Alicia Sit, Fr{\'e}d{\'e}ric Bouchard, Robert Fickler, J{\'e}r{\'e}mie
  Gagnon-Bischoff, Hugo Larocque, Khabat Heshami, Dominique Elser, Christian
  Peuntinger, Kevin G{\"u}nthner, Bettina Heim, et~al.
\newblock High-dimensional intra-city quantum cryptography with structured
  photons.
\newblock {\em arXiv preprint arXiv:1612.05195}, 2016.

\bibitem{Smania2016}
Massimiliano Smania, Ashraf~M Elhassan, Armin Tavakoli, and Mohamed Bourennane.
\newblock Experimental quantum multiparty communication protocols.
\newblock {\em Npj Quantum Information}, 2:16010, 2016.

\bibitem{Lee:2016vk}
Catherine Lee, Darius Bunandar, Zheshen Zhang, Gregory~R Steinbrecher, P~Ben
  Dixon, Franco N~C Wong, Jeffrey~H Shapiro, Scott~A Hamilton, and Dirk
  Englund.
\newblock {High-rate field demonstration of large-alphabet quantum key
  distribution}.
\newblock {\em arXiv}, November 2016.

\bibitem{vaziri2002experimental}
Alipasha Vaziri, Gregor Weihs, and Anton Zeilinger.
\newblock Experimental two-photon, three-dimensional entanglement for quantum
  communication.
\newblock {\em Physical Review Letters}, 89(24):240401, 2002.

\bibitem{cai2016new}
Yu~Cai, Jean-Daniel Bancal, Jacquiline Romero, and Valerio Scarani.
\newblock A new device-independent dimension witness and its experimental
  implementation.
\newblock {\em Journal of Physics A: Mathematical and Theoretical},
  49(30):305301, 2016.

\bibitem{allen1992orbital}
Les Allen, Marco~W Beijersbergen, RJC Spreeuw, and JP~Woerdman.
\newblock Orbital angular momentum of light and the transformation of
  laguerre-gaussian laser modes.
\newblock {\em Physical Review A}, 45(11):8185, 1992.

\bibitem{krenn2017orbital}
Mario Krenn, Mehul Malik, Manuel Erhard, and Anton Zeilinger.
\newblock Orbital angular momentum of photons and the entanglement of
  laguerre--gaussian modes.
\newblock {\em Phil. Trans. R. Soc. A}, 375(2087):20150442, 2017.

\bibitem{mair2001entanglement}
Alois Mair, Alipasha Vaziri, Gregor Weihs, and Anton Zeilinger.
\newblock Entanglement of the orbital angular momentum states of photons.
\newblock {\em Nature}, 412(6844):313--316, 2001.

\bibitem{Mirhosseini:2013go}
Mohammad Mirhosseini, Omar~S Maga{\~n}a-Loaiza, Changchen Chen, Brandon
  Rodenburg, Mehul Malik, and Robert~W Boyd.
\newblock {Rapid generation of light beams carrying orbital angular momentum}.
\newblock {\em Optics Express}, 21(25):30204--30211, December 2013.

\bibitem{malik2014direct}
Mehul Malik, Mohammad Mirhosseini, Martin~PJ Lavery, Jonathan Leach, Miles~J
  Padgett, and Robert~W Boyd.
\newblock Direct measurement of a 27-dimensional orbital-angular-momentum state
  vector.
\newblock {\em Nature communications}, 5, 2014.

\bibitem{krenn2016automated}
Mario Krenn, Mehul Malik, Robert Fickler, Radek Lapkiewicz, and Anton
  Zeilinger.
\newblock Automated search for new quantum experiments.
\newblock {\em Physical review letters}, 116(9):090405, 2016.

\bibitem{schlederer2016cyclic}
Florian Schlederer, Mario Krenn, Robert Fickler, Mehul Malik, and Anton
  Zeilinger.
\newblock Cyclic transformation of orbital angular momentum modes.
\newblock {\em New Journal of Physics}, 18(4):043019, 2016.

\bibitem{leach2002measuring}
Jonathan Leach, Miles~J Padgett, Stephen~M Barnett, Sonja Franke-Arnold, and
  Johannes Courtial.
\newblock Measuring the orbital angular momentum of a single photon.
\newblock {\em Physical review letters}, 88(25):257901, 2002.

\bibitem{de2005implementing}
AN~De~Oliveira, SP~Walborn, and CH~Monken.
\newblock Implementing the deutsch algorithm with polarization and transverse
  spatial modes.
\newblock {\em Journal of Optics B: Quantum and Semiclassical Optics},
  7(9):288, 2005.

\bibitem{asadian2016heisenberg}
Ali Asadian, Paul Erker, Marcus Huber, and Claude Kl{\"o}ckl.
\newblock Heisenberg-weyl observables: Bloch vectors in phase space.
\newblock {\em Physical Review A}, 94(1):010301, 2016.

\bibitem{reck1994experimental}
Michael Reck, Anton Zeilinger, Herbert~J Bernstein, and Philip Bertani.
\newblock Experimental realization of any discrete unitary operator.
\newblock {\em Physical Review Letters}, 73(1):58, 1994.

\bibitem{schaeff2015experimental}
Christoph Schaeff, Robert Polster, Marcus Huber, Sven Ramelow, and Anton
  Zeilinger.
\newblock Experimental access to higher-dimensional entangled quantum systems
  using integrated optics.
\newblock {\em Optica}, 2(6):523--529, 2015.

\bibitem{carolan2015universal}
Jacques Carolan, Christopher Harrold, Chris Sparrow, Enrique
  Mart{\'\i}n-L{\'o}pez, Nicholas~J Russell, Joshua~W Silverstone, Peter~J
  Shadbolt, Nobuyuki Matsuda, Manabu Oguma, Mikitaka Itoh, et~al.
\newblock Universal linear optics.
\newblock {\em Science}, 349(6249):711--716, 2015.

\bibitem{romero2012increasing}
J~Romero, D~Giovannini, S~Franke-Arnold, SM~Barnett, and MJ~Padgett.
\newblock Increasing the dimension in high-dimensional two-photon orbital
  angular momentum entanglement.
\newblock {\em Physical Review A}, 86(1):012334, 2012.

\bibitem{krenn2015twisted}
Mario Krenn, Johannes Handsteiner, Matthias Fink, Robert Fickler, and Anton
  Zeilinger.
\newblock Twisted photon entanglement through turbulent air across vienna.
\newblock {\em Proceedings of the National Academy of Sciences},
  112(46):14197--14201, 2015.

\bibitem{yokoyama2013ultra}
Shota Yokoyama, Ryuji Ukai, Seiji~C Armstrong, Chanond Sornphiphatphong,
  Toshiyuki Kaji, Shigenari Suzuki, Jun-ichi Yoshikawa, Hidehiro Yonezawa,
  Nicolas~C Menicucci, and Akira Furusawa.
\newblock Ultra-large-scale continuous-variable cluster states multiplexed in
  the time domain.
\newblock {\em Nature Photonics}, 7(12):982--986, 2013.

\bibitem{xie2015harnessing}
Zhenda Xie, Tian Zhong, Sajan Shrestha, XinAn Xu, Junlin Liang, Yan-Xiao Gong,
  Joshua~C Bienfang, Alessandro Restelli, Jeffrey~H Shapiro, Franco~NC Wong,
  et~al.
\newblock Harnessing high-dimensional hyperentanglement through a biphoton
  frequency comb.
\newblock {\em Nature Photonics}, 9(8):536--542, 2015.

\bibitem{muthukrishnan2000multivalued}
Ashok Muthukrishnan and CR~Stroud~Jr.
\newblock Multivalued logic gates for quantum computation.
\newblock {\em Physical Review A}, 62(5):052309, 2000.

\bibitem{klimov2003qutrit}
AB~Klimov, R~Guzman, JC~Retamal, and Carlos Saavedra.
\newblock Qutrit quantum computer with trapped ions.
\newblock {\em Physical Review A}, 67(6):062313, 2003.

\bibitem{smith2013quantum}
A~Smith, BE~Anderson, H~Sosa-Martinez, CA~Riofrio, Ivan~H Deutsch, and Poul~S
  Jessen.
\newblock Quantum control in the cs 6 s 1/2 ground manifold using
  radio-frequency and microwave magnetic fields.
\newblock {\em Physical review letters}, 111(17):170502, 2013.

\bibitem{hofheinz2009synthesizing}
Max Hofheinz, H~Wang, Markus Ansmann, Radoslaw~C Bialczak, Erik Lucero, Matthew
  Neeley, AD~O'Connell, Daniel Sank, J~Wenner, John~M Martinis, et~al.
\newblock Synthesizing arbitrary quantum states in a superconducting resonator.
\newblock {\em Nature}, 459(7246):546--549, 2009.

\bibitem{gottesman1999fault}
Daniel Gottesman.
\newblock Fault-tolerant quantum computation with higher-dimensional systems.
\newblock In {\em Quantum Computing and Quantum Communications}, pages
  302--313. Springer, 1999.

\bibitem{lawrence2004mutually}
Jay Lawrence.
\newblock Mutually unbiased bases and trinary operator sets for n qutrits.
\newblock {\em Physical Review A}, 70(1):012302, 2004.

\bibitem{agnew2013generation}
Megan Agnew, Jeff~Z Salvail, Jonathan Leach, and Robert~W Boyd.
\newblock Generation of orbital angular momentum bell states and their
  verification via accessible nonlinear witnesses.
\newblock {\em Physical review letters}, 111(3):030402, 2013.

\bibitem{wang2015quantum}
Xi-Lin Wang, Xin-Dong Cai, Zu-En Su, Ming-Cheng Chen, Dian Wu, Li~Li, Nai-Le
  Liu, Chao-Yang Lu, and Jian-Wei Pan.
\newblock Quantum teleportation of multiple degrees of freedom of a single
  photon.
\newblock {\em Nature}, 518(7540):516--519, 2015.

\bibitem{ionicioiu2016sorting}
Radu Ionicioiu.
\newblock Sorting quantum systems efficiently.
\newblock {\em Scientific reports}, 6, 2016.

\bibitem{10.1088/2058-9565/aa5917}
Manuel Erhard, Mehul Malik, and Anton Zeilinger.
\newblock A quantum router for high-dimensional entanglement.
\newblock {\em Quantum Science and Technology}, 2017.

\bibitem{Feiran}
Feiran Wang~et al.
\newblock Generation of the complete four-dimensional bell basis.
\newblock {\em (in preparation)}, 2016.

\bibitem{lapkiewicz2011experimental}
Radek Lapkiewicz, Peizhe Li, Christoph Schaeff, Nathan~K Langford, Sven
  Ramelow, Marcin Wie{\'s}niak, and Anton Zeilinger.
\newblock Experimental non-classicality of an indivisible quantum system.
\newblock {\em Nature}, 474(7352):490--493, 2011.

\bibitem{dada2011experimental}
Adetunmise~C Dada, Jonathan Leach, Gerald~S Buller, Miles~J Padgett, and Erika
  Andersson.
\newblock Experimental high-dimensional two-photon entanglement and violations
  of generalized bell inequalities.
\newblock {\em Nature Physics}, 7(9):677--680, 2011.

\bibitem{mafu2013higher}
Mhlambululi Mafu, Angela Dudley, Sandeep Goyal, Daniel Giovannini, Melanie
  McLaren, Miles~J Padgett, Thomas Konrad, Francesco Petruccione, Norbert
  L{\"u}tkenhaus, and Andrew Forbes.
\newblock Higher-dimensional orbital-angular-momentum-based quantum key
  distribution with mutually unbiased bases.
\newblock {\em Physical Review A}, 88(3):032305, 2013.

\bibitem{mirhosseini2015high}
Mohammad Mirhosseini, Omar~S Maga{\~n}a-Loaiza, Malcolm~N O’Sullivan, Brandon
  Rodenburg, Mehul Malik, Martin~PJ Lavery, Miles~J Padgett, Daniel~J Gauthier,
  and Robert~W Boyd.
\newblock High-dimensional quantum cryptography with twisted light.
\newblock {\em New Journal of Physics}, 17(3):033033, 2015.

\bibitem{hill2016hyperdense}
Alexander Hill, Trent Graham, and Paul Kwiat.
\newblock Hyperdense coding with single photons.
\newblock In {\em Frontiers in Optics}, pages FW2B--2. Optical Society of
  America, 2016.

\bibitem{ralph2007efficient}
TC~Ralph, KJ~Resch, and A~Gilchrist.
\newblock Efficient toffoli gates using qudits.
\newblock {\em Physical Review A}, 75(2):022313, 2007.

\bibitem{lanyon2009simplifying}
Benjamin~P Lanyon, Marco Barbieri, Marcelo~P Almeida, Thomas Jennewein,
  Timothy~C Ralph, Kevin~J Resch, Geoff~J Pryde, Jeremy~L O’brien, Alexei
  Gilchrist, and Andrew~G White.
\newblock Simplifying quantum logic using higher-dimensional hilbert spaces.
\newblock {\em Nature Physics}, 5(2):134--140, 2009.

\bibitem{bocharov2016factoring}
Alex Bocharov, Martin Roetteler, and Krysta~M Svore.
\newblock Factoring with qutrits: Shor's algorithm on ternary and metaplectic
  quantum architectures.
\newblock {\em arXiv preprint arXiv:1605.02756}, 2016.

\bibitem{garcia2013swap}
Juan~Carlos Garcia-Escartin and Pedro Chamorro-Posada.
\newblock A swap gate for qudits.
\newblock {\em Quantum information processing}, 12(12):3625--3631, 2013.

\end{thebibliography}
\section*{Appendix}
\subsection{All unitaries from X and Z}
Here we briefly show that it is possible to construct every unitary transformation in four dimensions in terms of X and Z-gates, and integer powers of them. Here we follow the construction in \cite{asadian2016heisenberg}. The derivation is general for arbitrary dimensions $d$. For the purpose of our experiment, in the end we will make $d \to 4$.

We start by defining the operator (called Heisenberg-Weyl operators)
\begin{equation}
\mathcal{D}(l,m)=\exp\left(i\frac{\pi l m}{2}\right) Z^l X^m. 
\end{equation}
From these operators, we can write a minimum and complete set of Hermitian operators
\begin{equation}
\mathcal{Q}_{l,m}=\chi \mathcal{D}(l,m) + \chi^* \mathcal{D}(l,m)^{\dagger}, 
\end{equation}
with $\chi=\frac{1+i}{2}$. With a linear superposition of the basis elements, arbitrary hermitian matrices can be constructed in the form
\begin{equation}
A=\sum_{l=0}^{d-1} \sum_{m=0}^{d-1} c_{l,m} \mathcal{Q}_{l,m}, 
\end{equation}
with real coefficients $c_{l,m}$, and $d$ being the dimension of the Hilbert space. Every unitary transformation can be written as the exponent of a hermitian generator, such that
\begin{equation}
U=\exp\left(i A \right)=\sum_{n=0}^{\infty}\frac{i^n}{n!}A^n. 
\end{equation}
U has infinitely many terms which are combinations of X and Z gates, and X and Z do not commute. However, because of $X\cdot Z = - i Z X$, one can always write
\begin{equation}
U=\sum_{l=0}^{d-1} \sum_{m=0}^{d-1} g_{l,m} X^l Z^m,
\end{equation}
with complex coefficients $g_{l,m}$. For $d=4$, due to $X^4=1$ and $Z^4=1$, the sum only goes from $0 \leq l,m \leq 3$. Therefore, one arrives at
\begin{equation}
U=\sum_{l=0}^{3} \sum_{m=0}^{3} h_{l,m} X^l Z^m,
\end{equation}
with complex coefficients $h_{l,m}$. This sum has 16 terms with various coefficients. That means, in principle one could apply the 16 different operations in superposition and create every arbitrary unitary transformation in a 4-dimensional space. While this is experimentally challenging, in many important cases, the formula simplifies significantly.
\end{document}